# Small brains but big challenges:
# White matter tractography in early life samples

Jessica Dubois[1,2]*, Mareike Grotheer[3,4], Joseph Yuan-Mou Yang[5,6,7], Jacques-Donald Tournier[8,9], Christian Beaulieu[10], Catherine Lebel[11]

**Affiliations**

1. Université Paris Cité, Inserm, NeuroDiderot, F-75019 Paris, France

2. Université Paris-Saclay, CEA, NeuroSpin, UNIACT, F-91191, Gif-sur-Yvette, France

3. Department of Psychology, Phillips-Universität Marburg, Marburg, 35039, Germany

4. Center for Mind, Brain and Behavior, Phillips-Universität Marburg, Justus-Liebig Universität Giessen, and Technische Universität Darmstadt, Marburg, 35039, Germany

5. Department of Neurosurgery, Neuroscience Advanced Clinical Imaging Service (NACIS), The Royal Children's Hospital, Melbourne, Australia

6. Neuroscience Research, Murdoch Children's Research Institute, Melbourne, Australia

7. Department of Paediatrics, University of Melbourne, Melbourne, Australia

8. Department of Biomedical Engineering, School of Biomedical Engineering and Imaging Sciences, King's College London, King's Health Partners, St. Thomas' Hospital, London, UK

9. Centre for the Developing Brain, School of Biomedical Engineering and Imaging Sciences, King's College London, King's Health Partners, St. Thomas' Hospital, London, UK

10. Departments of Radiology and Diagnostic Imaging & Biomedical Engineering, University of Alberta, Edmonton, Alberta, Canada

11. Department of Radiology, University of Calgary

* Corresponding author: jessica.dubois@inserm.fr

**Abstract**

In the human brain, white matter development is a complex and long-lasting process involving intermingling micro- and macrostructural mechanisms, such as fiber growth, pruning and myelination. Did you know that all these neurodevelopmental changes strongly affect MRI signals, with consequences on tractography performances and reliability? This communication aims to elaborate on these aspects, highlighting the importance of tracking and studying the developing connections with dedicated approaches.

**Introduction**

Understanding how networks develop in the human brain is a crucial question for comprehending how infants acquire remarkable abilities such as language or social cognition [1]. It is also essential for evaluating early disturbances that might lead to neurodevelopmental disorders originating as early as the prenatal period. Therefore, the ability to map and assess the development of these networks in a reliable, quantitative, and reproducible manner using non-invasive methods has been at the core of research for the past two decades. In the adult brain, functional networks are organized around specialized regions that are distributed across the brain and communicate with each other via white matter connections. Diffusion MRI combined with tractography has become the method of choice for *in vivo* mapping of these structural connections [2].

However, did you know that these connections begin to form as early as the first trimester of gestation, enabling the emergence of brain activity, and that their development is not complete until early adulthood for some functional systems? While age-related diffusion MRI metric curves show extrema between 20 and 40 years of age depending on white matter pathways, it's during the first two post-natal years that the changes are most intense. This is because the fetal, neonatal, and early childhood periods are particularly rich in white matter changes, in addition to dramatic brain growth [3, 4]. Then, did you know that applying diffusion-related tractography techniques - originally developed for the adult brain - to the early developing brain presents numerous challenges? On top of this, the tracts in newborns and infants are very small, and experimental constraints often limit acquisition time and therefore accessible spatial resolution and data quality (i.e., without motion artifacts).

In complement to extensive methodological reviews on diffusion MRI and tractography in infants (e.g., [5, 6]), the aim of this communication is to focus on the key mechanisms underlying the development of white matter connections (Figure 1a) and to draw attention to how this might impact tractography while introducing potential approaches to address these challenges.

**Connection growth and pruning**

Neuronal axons start to grow during the early fetal period, at around 10 weeks of gestational age (w GA) [3]. The first "pioneering" axons are guided by their extremity growth cones to reach their target structure and others follow through the process of "fasciculation" [7]. These growing connections constitute the intermediate zone that will later be called the white matter. The intermediate zone lies between central structures (proliferative zones, central grey nuclei) and the peripheral developing cortical plate and subplate. This latter structure is an important transitory compartment for the "waiting" of afferent fibers (e.g. thalamo-cortical connections) and the creation of efferent fibers (e.g. cortico-cortical connections) [3]. In the midfetal period, axonal growth continues, and efferent fibers extend to their targets in the striatum, pons, and spinal cord. Towards the end of gestation, when preterm born neonates are already viable, afferent fibers from the subplate reallocate to establish permanent connectivity with cortical neurons, and the brain connectivity is very rich. Globally, early signs of growth / complete axonal growth have been observed at different ages for different white matter pathways: projection fibers 10-12 / 26-34 w GA; commissural fibers 13-15 / 28-34 w GA; association fibers 19-21 / 35-37 w GA [8]. At full-term birth, major long-distance pathways are thus supposed to be in place. The initial production stage is followed by a pruning stage characterized by suppression of redundant or aberrant circuits, which is dramatically sensitive to the environment, especially during the first postnatal year particularly for callosal axons [3]. U-fibers further develop, and cortico-cortical connectivity is reorganized by synaptogenesis and pruning, growth of dendritic arborization, etc.

An interesting example of developing bundles is the cortico-spinal tract, which shows an initial prenatal growth of connections contralaterally and ipsilaterally from the primary motor cortex of each hemisphere. During normal development, ipsilateral connections are gradually withdrawn under the influence of inter-connection competition mechanisms, giving rise to the well-known organization of the sensorimotor system in adults (right hemisphere controlling the left side of the body, and vice versa) (Figure 1b). However, in certain pathological cases, these ipsilateral fibers do not regress: if one of the two motor brain regions is damaged at an early stage (e.g., in perinatal stroke), the contralateral body part will continue to receive connections from the uninjured hemisphere, being "controlled" by it [9]. This might give rise to altered "reorganization" and involuntary mirror movements of paretic and non-paretic hands in children with cerebral palsy.

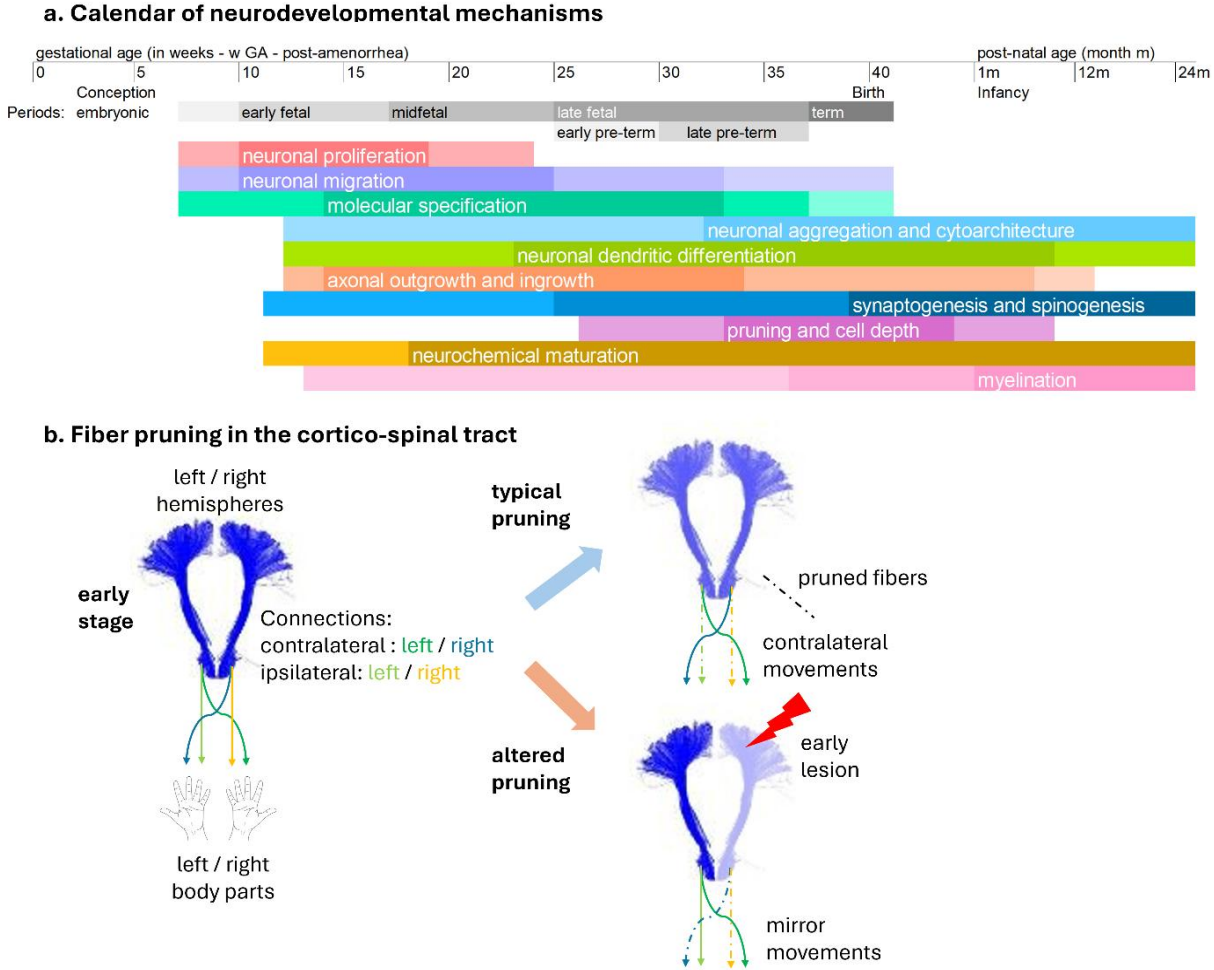

Figure 1: a. Calendar of developmental periods in the first 1000 days, with schematic representation of key neurogenetic mechanisms, inspired by [3]. b. Schematic representation of fiber pruning progression in the developing cortico-spinal tract, inspired by [9]. At an early stage (left column), connections from the two primary motor cortical regions grow both contralaterally and ipsilaterally. During normal development (right column, upper row), pruning is observed for ipsilateral connections from both hemispheres, allowing contralateral movements. But in case of early lesion in one hemisphere (right column, upper row), fibers from that side are pruned, while fibers from the opposite hemisphere are preserved: the uninjured hemisphere then controls the two sides of the body, giving rise to mirror movements.

**Fiber myelination**

In parallel to axonal growth and pruning, myelination plays an important role in white matter development and maturation. It is based on the proliferation of oligodendrocyte precursors, their migration along axons, and the formation of the myelin sheath that increases the conduction speed of neural information [10]. Myelination further contributes to the stabilization of connections by preventing them from pruning. With a peak during the first post-natal year, myelination progresses asynchronously across cerebral regions and functional systems, with a caudo-rostral gradient and earlier maturation of projection sensory and motor pathways compared to associative ones.

Myelination and other microstructural developmental mechanisms (e.g. proliferation of glial cells, development of dendritic arborizations in grey matter) are associated with significant changes in brain tissue composition and macrostructure during early development: water content decreases while lipid and iron content increases, volumes of grey and white matter significantly increase, etc. All this has a major impact on the signals measured with MRI techniques: T1 and T2 relaxation times decrease, as does the apparent diffusion coefficient. These effects differ across tissues with different compositions, affecting image contrast mainly during the first postnatal year [4]. T1- and T2-weighted images then show opposite contrasts between grey and white matter in newborns compared to adults (Figure 2a), and a loss of grey-white contrast is observed between 6 months and 1 year of age. The age-related decrease in T2 relaxation times also indirectly impacts diffusion MRI information since it decreases signal-to-noise ratio and can bias the estimation of metrics derived from diffusion models (e.g. diffusion tensor imaging). In addition, diffusion metrics show strong changes through the course of white matter development likely because of myelination as well as axon packing within tracts [11], and this proceeds with different maturation calendars across pathways [12, 13].

**a. T2w images at different ages**

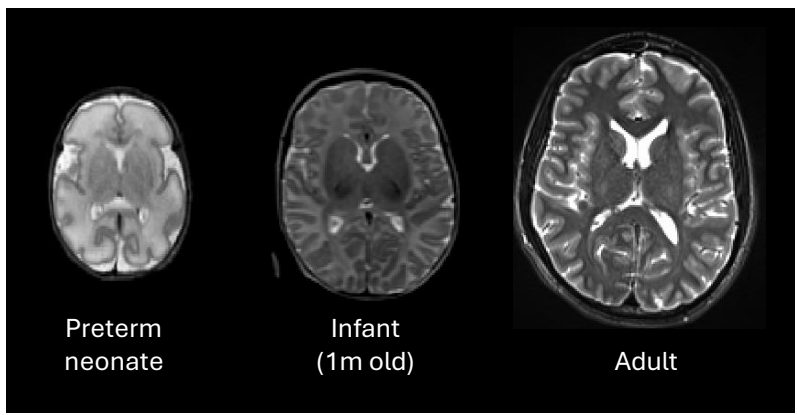

**b. Examples of tractography reconstructions**

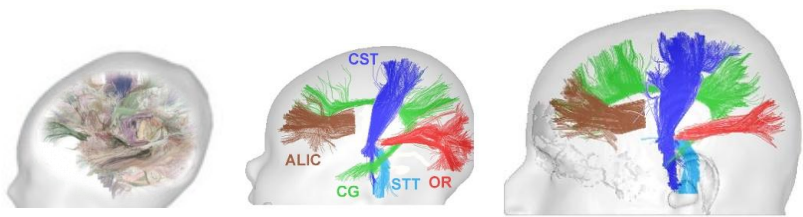

*Figure 2: Illustration of (a) the changing contrast in T2-weighted (T2w) MRI images over development and (b) tractography reconstructions at different ages (left column: preterm neonate, middle column: infant, right column: adult).*

**Challenges of tractography in the developing white matter**

Given this complex interplay of neurodevelopmental mechanisms, identifying and investigating the developing pathways using diffusion MRI and tractography is a difficult task [5, 6]. First, it is worth noting that the tractography reliability highly depends on diffusion MRI data quality (signal-to-noise ratio, spatial resolution, b-value(s), number of directions), the considered diffusion model, the tractography algorithm, the parameter tuning, etc. [6]. Several tractography challenges are due to the previously described neurobiological properties and constraints of the developing brain, in particular the low myelination and axon packing/coherence of white matter fascicles leading to low anisotropy values [11]. Even though thresholding anisotropy values is not essential for defining tractography masks, the reliability and performance of most tracking algorithms remain influenced by fascicle characteristics including myelination. Therefore, this could bias the reconstruction (and quantification) of pathways at different maturation stages (i.e., a given pathway at different ages or different pathways at a given age).

In addition, brain structures and white matter pathways are very small at birth, making the image spatial resolution a major issue in babies and young children while acquisition time is limited. For images of similar spatial resolution, partial volume effects are larger in small brains compared to adult brains, and this also contributes to lower anisotropy values in small tracts. The effective resolution (i.e. relative to the size of the growing structures) improves over the course of development. This provides challenges for interpreting fiber directions in crossing regions comparably, especially when crossing bundles mature at different rates. Overall, disentangling axonal growth, pruning and myelination in a whole-brain set of tracked fibers is not possible with current methods.

Further, the quality of diffusion MRI data that can be reasonably obtained in neonates and infants is limited by the time available for acquisition and by potential movement artifacts. Quality depends on the acquisition settings and the clinical versus research context (e.g., possibility of sedation or natural sleep). One of today's most advanced protocols for neonatal brain is the one from the developing Human Connectome Project, achieving 1.5mm isotropic resolution but with 20min acquisition time [14].

Finally, tractography protocols often rely on the processing of anatomical images for the definition of regions-of-interest (i.e., seeds, waypoints and endpoints), propagation mask, etc. As this requires tissue segmentation and parcellation, the changing contrast of anatomical T1-weighted or T2-weighted images could impact reproducibility across infants and children at different ages. The difficulty to discern the boundary between grey and white matter is a real problem for techniques like anatomically constrained tractography, particularly in the early stages as the anisotropy in the cortex is often higher than in the subcortical white matter. Preventing streamlines from tracking through cortex into adjacent gyri is highly challenging.

**Potential solutions to face tractography challenges**

Despite these challenges, diffusion MRI allows the identification of main white matter pathways, and their 3D mapping with tractography is feasible from the full-term period (Figure 2b), at least for long distance projection, commissural, limbic, and association tracts [13, 15, 16]. At the data acquisition level, while consensus on optimal protocols is missing, successful tractography reconstructions have been obtained with complex diffusion protocols (e.g., multi-shell protocol of the developing Human Connectome Project [14, 17, 18]) but also with simpler (and faster) scan parameters (e.g., [19]). Nowadays, growing interest in the development of short-distance connections (e.g., U- fibers) is driving the use of acquisition protocols with higher spatial resolution (<1.5mm) and minimizing partial volume effects with cerebrospinal fluid, which is also advantageous for separating tracts in complex areas with crossings. The quality of diffusion imaging and derived tractography, as well as acquisition speed, will also potentially improve in the coming years given advances in MRI technology (higher static magnetic fields, stronger gradients, sensitive phased array RF coils, etc.).

In terms of data post-processing, most pediatric studies have historically used the diffusion tensor model for tractography. While this provides useful information, it is limited, and it is now important to go beyond: using higher order models have become fairly common for tractography. Based on high angular resolution diffusion data and multi-tissue constrained spherical deconvolution technique, considering several anisotropic compartments is also an interesting direction to explore to better describe the fiber directions in the developing white matter [20]. But it opens many questions about interpretation and how to perform tractography based on such information. Regarding tractography per se, the use of deterministic algorithms with regularization [19] or probabilistic algorithms [17, 18] have been proposed to better approximate the fiber pathways.

For a given model and tractography setting, different approaches have further been proposed to identify tracts in neonates in a systematic and automated way through the robust definition of seeds, waypoints and endpoints [16, 17]. Informing tractography reconstruction by the fascicle microstructure [21] might also help for reconstructing connections at different maturation stages, and disentangling axonal growth, pruning and myelination processes. In the end, it's worth asking about the potential of considering longitudinal datasets - when available! - to better account for the inter-individual variability: informing tractography reconstructions at young ages by those at later ages could improve the reliability of tract quantifications (e.g., number of fibers, diffusion metrics), but this remains to be tested.

**Conclusion**

Tractography approaches have immense potential for studying connections in the developing brain. Yet there are still many advances to be made to explore subtle questions, such as the role of certain connections in the development of complex cognitive learning processes, or the impact of under- or over-connectivity phenomena in the emergence of neurodevelopmental disorders. Hence the developing brain provides all the ingredients for interesting discussions to come between neuroanatomists, developmentalists and tracto-methodologists.

**Funding sources:**

JD acknowledges funding support from the French government as part of the France 2030 programme (grant ANR-23-IAIIU-0010, IHU Robert-Debré du Cerveau de l'Enfant), the French National Agency for Research (grant ANR-22-CE37-0028), the IdEx Université de Paris (ANR-18-IDEX-0001), the Fondation de France and Fondation Médisite (grants FdF-18-00092867 and FdF-20-00111908). Dr JYMY acknowledges position funding support from the Royal Children's Hospital Foundation (RCHF 2022–1402), and support from The Kids' Cancer Project (TKCP) Col Reynolds Fellowship.